\documentclass[a4paper]{jpconf}
\usepackage{graphicx}
\begin{document}
\title{Nuclear masses, deformations and shell effects}

\author{Jorge G. Hirsch }

\address{Instituto de Ciencias
    Nucleares, Universidad Nacional Aut\'{o}noma de M\'{e}xico, 04510
    M\'{e}xico, D.F., Mexico}

\ead{hirsch@nucleares.unam.mx}

\author{C\'esar Barbero and Alejandro E. Mariano }

\address{Departamento de F\'\i sica, Facultad de Ciencias Exactas, Universidad Nacional de La Plata,
CC 67, 1900 La Plata, Argentina}

\begin{abstract}
We show that the Liquid Drop Model is best suited to describe the masses of prolate deformed nuclei than of spherical nuclei.
To this end three Liquid Drop Mass formulas are employed to describe nuclear masses of eight sets of nuclei with similar quadrupole deformations. It is shown that they are able to fit the measured masses of prolate deformed nuclei with an RMS smaller than 750 keV, while for the spherical nuclei the RMS is, in the three cases, larger than 2000 keV. The RMS of the best fit of the masses of semi-magic nuclei is also larger than 2000 keV. The parameters of the three models are studied, showing that the surface symmetry term is the one which varies the most from one group of nuclei to another. In one model, isospin dependent terms are also found to exhibit strong changes. The inclusion of shell effects allows for better fits,  which continue to be better in the prolate deformed nuclei region.    
\end{abstract}

\section{Introduction}

The description of nuclear masses in terms of the Liquid Drop Model paved the way to the basic understanding of nuclear properties, like the saturation of the nuclear force, the existence of pairing and shell effects, and the description of fission and fusion processes \cite{Bohr}. The Q-values of different nuclear reactions, obtained from mass differences, must be accurately known to allow the description of the astrophysical origin of the elements \cite{Rol88}. Accurate theoretical predictions of nuclear masses remain a challenge \cite{Bla06}, sharing the difficulties with other quantum many-body calculations, and complicated by the absence of a full theory of the nuclear interaction.  

Decades of work have produced microscopic and macroscopic mass formulas \cite{Lunn03}. At present, the most successful approaches seem to be the Finite Range Droplet Model (FRDM) \cite{Mol95}, the Skyrme and Gogny Hartee Fock Bogolyubov (HFB) \cite{Gor01,Gor09}, and the Duflo-Zuker (DZ) mass formula \cite{Duf94,Zuk94,Duf95}. They allow for the calculation of masses, charge radii, deformations, and in some cases also fission barriers. They all contain a macroscopic sector which resembles the Liquid Drop Mass (LDM) formula, and include deformation effects. 
HFB calculations are now able to fit known nuclear masses with deviations competitive with the '95 FRDM calculations, which are also being improved, while the most precise and robust nuclear mass predictions are given by the DZ model \cite{Lunn03,Men08}.

The Liquid Drop Mass (LDM) formula captures the macroscopic features of the mass dependence on the number of neutrons $N$, of protons $Z$, and on its mass numbers $A=N+Z$. It  includes volume and surface terms, the Coulomb interaction between protons, Wigner and symmetry terms, linear and quadratic in the neutron excess $N-Z$, and a pairing term. It is generally assumed that the liquid-drop energy of a spherical nucleus is described by a Bethe-Weizs\"acker mass formula \cite{Wang10}, being a common practice to describe nuclear masses and radii of spherical closed-shell nuclei in terms of a mean field and add deformation and other shell effects as corrections \cite{Die09}.

It is the purpose of this contribution to show that, when nuclei with measured masses are grouped according with their quadrupole deformations, three Liquid Drop Mass formulas consistently allow for a fit with an RMS smaller than 750 keV for the set of most prolate deformed nuclei, while they are unable to fit the masses of spherical nuclei with an RMS smaller than 2000 keV. Semi-magic nuclei are fit with a similarly large RMS.
The parameters of the three models are studied, showing that the surface symmetry term is the one which varies the most from one group of nuclei to another. In one model, isospin dependent terms are also found to exhibit strong changes. 

Shell effects refer to the differences between the experimental binding energies \cite{AME03} and the LDM predictions.  In this work they are included employing linear, quadratic \cite{Die07,JMT}, 3- and 4-body terms  \cite{Die09}, functions of the number of valence nucleons,  following the ideas of the Duflo-Zuker model \cite{Duf94,Zuk94,Duf95}.The inclusion of shell effects allows for better fits, while the smallest rms are found again in the prolate deformed nuclei region.  

\section{The three Liquid Drop Mass formulas}

Three Liquid Drop Mass formulas will be employed to analyze their ability to fit nuclear masses. One of them  treats consistently volume and surface effects \cite{Die09}. The second one incorporates explicitly isospin effects  \cite{Wang10} while the third one  includes quadratic isospin effects, the diffuseness correction to the Coulomb energy, the charge exchange correction term and the curvature energy \cite{Roger10}.

In Ref. \cite{Die09}  an  improved version of the liquid-drop mass formula with modified symmetry and Coulomb terms is built, following  a consistent treatment of nuclear bulk and surface effects. The negative nuclear interaction energy is given by:

\begin{equation}
E_{LDM1} = -a_v  A + a_s  A^{2/3} 
+ S_v \frac {4 T(T+1)}{A(1+yA^{-1/3})}
+ a_c \frac{Z (Z-1)}{(1-\Lambda) A^{1/3}} 
- a_p \frac {\Delta}{A^{1/3}},
\label{piet}
\end{equation}
where the pairing interaction is given by $\Delta$ = 2, 1, and 0 for even-even, odd-mass and odd-odd nuclei, respectively. 
A modification $\Lambda$ in the Coulomb term is included 
\begin{equation}
\Lambda = \frac {N - Z} {6Z(1 + y^{-1}A^{1/3})}
- \frac {5 \pi^2} {6} 
\frac {d^2} {r_0^2 A^{2/3}} ,
\end{equation}
Its presence in the denominator of the Coulomb term suggests that it can be viewed as a correction to
the radius of the nucleus. The symmetry term employs $4T(T + 1)$ instead of $(N - Z)^2$ to account for the Wigner energy. The Coulomb interaction is proportional to  $Z(Z - 1)$ to avoid the Coulomb interaction of a proton with itself.

The second LDM formula we are going to analyze was introduced in Ref. \cite{Wang10} by considering isospin effects.  
In this case the liquid-drop energy of a spherical nucleus is described by a modified Bethe-Weizs\"acker mass formula 
\begin{equation}
E_{LDM2} = -a_v  A + a_s  A^{2/3} 
+ a_{sym} I^2 A 
+ a_c \frac{Z (Z-1)}{A^{1/3}} (1- Z^{2/3}) 
- a_{pair} \frac {\delta_{np}}{A^{1/3}}, 
\label{wang}
\end{equation}
with isospin asymmetry $I = (N - Z)/A$. The pairing term is taken form \cite{Men10}
\begin{equation}
\delta_{np} = \left\{
\begin{array}{ll}
2 - \mid I \mid &\hbox{ :N and Z even}\\
\mid I \mid &\hbox{ :N and Z odd}\\
1 -\mid I \mid &\hbox{ :N even, Z odd, and N$>$Z}\\
1 - \mid I \mid &\hbox{ :N odd, Z even, and N$<$Z}\\
1 &\hbox{ :N even, Z odd, and N$<$Z}\\
1 &\hbox{ :N odd, Z even, and N$>$Z.}
\end{array}
\right.
\end{equation}
and the symmetry energy coefficient of finite nuclei is written as,
\begin{equation}
a_{sym} = c_{sym} \left[ 1 - \frac {\kappa} {A^{1/3}}  + \frac {2 - \mid I \mid} {2 + \mid I \mid A} \right]
\end{equation}
based on the conventional surface-symmetry term of liquid-drop model, with a small correction term for description of isospin dependence of $a_{sym}$. The symmetry energy coefficient $a_{sym}$ increases with decreasing isospin asymmetry. This $I$ correction term approximately describes the Wigner effect for heavy nuclei.

Different mass formulae derived from the liquid drop model and the pairing and shell energies of the
Thomas-Fermi model have been studied and compared in Ref \cite{Roger10}.
We selected for this study to include the diffuseness correction to the Coulomb energy, the charge exchange correction term and the curvature energy. In Ref. \cite{Roger10} it is reported that the
Coulomb diffuseness correction $Z^2/A$ term and the charge exchange correction $Z^{4/3}/A^{1/3}$ term play the main role to improve the accuracy of the mass formula. The Wigner term and the curvature energy can
also be used separately but their coefficients are very unstable. 

Their LDM formula is
\begin{equation}
\begin{array}{rl}
E_{LDM3}(N,Z) =& -a_v (1 - k_v I^2) A + a_s  (1 - k_s I^2) A^{2/3} + a_k  (1 - k_k I^2) A^{1/3}
+ \frac 3 5 \frac{e^2 Z^2}{r_0 A^{1/3}}  
\\&- f_p Z^2/A  - a_{c,exc} Z^{4/3}/A^{1/3} 
+ E_{pair} 
\end{array}
\label{roger}
\end{equation}
The volume energy corresponding to the saturated exchange force and infinite nuclear matter is
given by the first term. $I^2 A$ is the asymmetry energy of the Bethe-Weizs\"acker mass formula.
The second term is the surface energy. Its origin is the deficit of binding energy of the nucleons
at the nuclear surface and corresponds to semi-infinite nuclear matter. The following term is the
curvature energy. It results from non-uniform properties which correct the surface energy and
depends on the mean local curvature. The decrease of binding energy due to the Coulomb repulsion is given by the fourth term, which has and adjustable charge radii $r_0 A^{1/3}$. The $Z^2/A$ term is the diffuseness correction to the sharp radius Coulomb energy, also called also the proton form-factor. The
$Z^{4/3}/A^{1/3}$ term is the charge exchange correction term. The pairing energies of the
Thomas-Fermi model \cite{Myers96} were employed.

\section{The fits}

The coefficients of the three LDM were selected to minimize the root mean square
deviation (RMS) when the predicted binding energies $BE_{\rm th}(N,Z)$
are compared with the experimental ones $BE_{\rm exp}(N,Z)$, reported
in AME03 \cite{AME03}, modified so as to include more realistically the electron
binding energies as explained in Appendix A of Lunney, Pearson and
Thibault~\cite{Lunn03}. 

\begin{equation}
{\rm RMS}=\left\{\frac{{\sum\left[BE_{\rm exp}(N,Z)-BE_{\rm
        th}(N,Z)\right]^2}}{N_{nucl}}\right\}^{1/2}.
\end{equation}
$N_{nucl}$ is the number of nuclei in each group, listed in the fourth row of Table \ref{def}. The minimization
procedure uses the routine Minuit \cite{Minuit}.

The fits were performed employing the masses of nine groups of nuclei: 

\begin{itemize}
\item all nuclei whose measured masses are reported in AME03 \cite{AME03}, which have $N,\,Z \geq 8$,
\item seven groups of nuclei whose quadrupole deformations, taken form the FRDM \cite{Moll95}, lie in the ranges listed in the second and third row of Table \ref{def}, 
\item the group of all semi-magic nuclei, having Z= 8, 20, 28, 50, 82 or  N=   8, 20, 28, 50, 82 or 126.
\end{itemize}
\begin{table}
\begin{tabular}{c||c|ccccccc|c}
\br
group & all  &	1	&2	&3	&4	&5	&6	&7	&semi-magic\\
\mr
$e_2$ min & -0.65 & -0.65  & -0.11 &	0.00 &	0.04 & 0.12 & 0.18 & 0.23 &\\
$e_2$ max &	0.65 & -0.11 & 0.00  & 0.04 & 0.12 & 0.18 & 0.23 &  0.65 &\\
$N_{nucl}$  &2149 & 258	& 252  &332	       &272       & 307 & 364  & 364   &185\\
\mr
\end{tabular}
\label{def}
\caption{The nine groups on nuclei employed in the present study, their range of quadrupole deformation, and their number of nuclei.}
\end{table}
Notice that group 1 contains most of the oblate nuclei, that the more spherical nuclei belong to group 3, and that the more prolate deformed nuclei are included in groups 6  and 7.

For each LDM equation, nine fits were performed, one for each group of nuclei. In this way, nine sets of parameters were obtained, which minimize the RMS of each group of nuclei. Employing these nine sets of parameters, the RMS were estimated for all groups, whose RMS are shown in the next subsections.

\subsection{Analysis of the LDM1 formula}

Here we show the results obtained using Eq. (\ref{piet})  for the nine regions.
In our calculations we select nine sets of fixed values for the parameters $a_v$, $a_s$, $a_c$,
$a_p$, $S_v$ and $y$.
The results are exhibited in Tables \ref{tab1} and \ref{tab2}.

\begin{table}[h]
\begin{center}
\caption {RMS (in keV) for the nine groups (columns), for the nine sets of parameters (rows), employing Eq. (\ref{piet})}
\label{tab1}
\bigskip
\begin{tabular}{c|c|ccccccc|c}
\hline \hline
       & all  &	1	&2	&3	&4	&5	&6	&7	&semi-magic\\
\hline
set$_{all}$&${\bf 2387}$&$1842$&$3814$&$3080$&$2037$&$1495$&$1815$&$2060$&$3888$\\ \mr
set$_1$&$3300$&${\bf 1313}$&$3113$&$3420$&$3277$&$3173$&$5194$&$1476$&$4016$\\
set$_2$&$4053$&$2917$&${\bf 1676}$&$2433$&$3088$&$4273$&$6243$&$4701$&$2570$\\
set$_3$&$3404$&$3313$&$2298$&${\bf 2063}$&$2556$&$3408$&$4016$&$4721$&$2600$\\
set$_4$&$2567$&$2431$&$3867$&$3096$&${\bf 1746}$&$1741$&$1975$&$2626$&$3901$\\
set$_5$&$2630$&$2097$&$4848$&$3696$&$2072$&${\bf 1053}$&$1360$&$1735$&$4521$\\
set$_6$&$2828$&$1942$&$5205$&$4262$&$2578$&$1469$&${\bf 870}$&$1296$&$5078$\\
set$_7$&$3169$&$1655$&$4077$&$4220$&$3543$&$2753$&$3643$&${\bf 746}$&$5016$\\ \mr
set$_{semi}$&$4514$&$3816$&$1930$&$2665$&$3316$&$4777$&$6729$&$5306$&${\bf 2113}$\\
\hline\hline \end{tabular} \end{center}
\end{table}

\begin{table}[h]
\begin{center}
\caption {Sets of parameters which minimize the RMS for the nine groups of nuclei, employing Eq. (\ref{piet}). In the last three rows the average value of each parameter, their dispersion and percentage variation $\% \hbox{var}=100 \frac {\sigma}{\mid \hbox{average} \mid }$ are shown.}
\label{tab2}
\bigskip
\begin{tabular}{c|cccccc}
\hline \hline
&$a_v$&$a_s$&$a_c$&$a_p$&$S_v$&$y$\\
\hline
set$_{all}$&$15.822$&$18.491$&$0.703$&$6.048$&$30.701$&$2.616$\\
set$_1$&$15.515$&$17.577$&$0.686$&$6.092$&$25.997$&$1.583$\\
set$_2$&$15.883$&$18.622$&$0.712$&$5.733$&$28.711$&$2.284$\\
set$_3$&$15.978$&$18.884$&$0.716$&$5.895$&$32.091$&$3.059$\\
set$_4$&$15.879$&$18.564$&$0.708$&$4.103$&$31.459$&$2.666$\\
set$_5$&$15.832$&$18.435$&$0.707$&$5.266$&$30.575$&$2.482$\\
set$_6$&$15.609$&$17.835$&$0.689$&$5.334$&$29.762$&$2.388$\\
set$_7$&$15.553$&$17.782$&$0.684$&$5.126$&$27.402$&$1.865$\\
set$_{semi}$&$15.775$&$18.012$&$0.712$&$5.500$&$26.407$&$1.489$\\ \mr
average&$15.761$&$18.245$&$0.702$&$5.455$&$29.234$&$2.270$\\
$\sigma$&$0.153$&$0.426$&$0.012$&$0.579$&$2.093$&$0.495$\\
$\%$ var &1.0        & 2.3        &  1.7     &  10.6   &    7.1       &    21.8 \\
\hline\hline \end{tabular} \end{center}
\end{table}

\subsection{Analysis of the LDM2 formula}

Here we show the results obtained using Eq. (\ref{wang})  for the nine regions.
In our calculations we select nine sets of fixed values for the parameters $a_v$, $a_s$, $a_c$, $a_{pair}$, $c_{sym}$ and $\kappa$. The results are exhibited in Tables \ref{tab3} and \ref{tab4}.

\begin{table}[h]
\begin{center}
\caption {RMS (in keV) for the nine groups (columns), for the nine sets of parameters (rows), employing Eq. (\ref{wang})}
\label{tab3}
\bigskip
\begin{tabular}{c|c|ccccccc|c}
\hline \hline
       & all  &	1	&2	&3	&4	&5	&6	&7	&semi-magic\\\hline
set$_{all}$&${\bf 2374}$&$1788$&$3733$&$3081$&$2044$&$1502$&$1855$&$2058$&$3916$\\ \mr
set$_1$&$3208$&${\bf 1254}$&$3114$&$3400$&$3243$&$3077$&$4947$&$1435$&$4060$\\
set$_2$&$4092$&$2965$&${\bf 1675}$&$2380$&$3027$&$4318$&$6393$&$4697$&$2380$\\
set$_3$&$3475$&$3352$&$2173$&${\bf 2069}$&$2596$&$3518$&$4183$&$4816$&$2551$\\
set$_4$&$2561$&$2408$&$3761$&$3052$&${\bf 1762}$&$1771$&$2036$&$2688$&$3897$\\
set$_5$&$2616$&$2073$&$4823$&$3702$&$2083$&${\bf 1021}$&$1311$&$1704$&$4548$\\
set$_6$&$2795$&$1865$&$5130$&$4234$&$2619$&$1435$&${\bf 838}$&$1242$&$5090$\\
set$_7$&$3074$&$1589$&$4216$&$4238$&$3541$&$2590$&$3138$&${\bf 656}$&$5093$\\ \mr
set$_{semi}$&$4677$&$3915$&$1959$&$2722$&$3427$&$4953$&$7036$&$5466$&${\bf 2056}$\\
\hline\hline \end{tabular} \end{center}
\end{table}

\begin{table}[h]
\begin{center}
\caption {Sets of parameters which minimize the RMS for the nine groups of nuclei, employing Eq. (\ref{wang}). In the last three rows the average value of each parameter, their dispersion and percentage variation $\% \hbox{var}=100 \frac {\sigma}{\mid \hbox{average} \mid }$ are shown.}
\label{tab4}
\bigskip
\begin{tabular}{c|cccccc}
\hline \hline
&$a_v$&$a_s$&$a_c$&$a_{pair}$&$c_{sym}$&$\kappa$\\
\hline
set$_{all}$&$15.711$&$18.920$&$0.720$&$6.989$&$30.045$&$1.587$\\
set$_1$&$15.432$&$18.119$&$0.700$&$6.793$&$26.488$&$1.260$\\
set$_2$&$15.695$&$18.802$&$0.722$&$7.443$&$27.979$&$1.441$\\
set$_3$&$15.903$&$19.395$&$0.737$&$6.717$&$30.582$&$1.642$\\
set$_4$&$15.729$&$18.877$&$0.721$&$4.685$&$30.858$&$1.635$\\
set$_5$&$15.685$&$18.756$&$0.721$&$6.210$&$29.969$&$1.548$\\
set$_6$&$15.493$&$18.254$&$0.703$&$6.276$&$29.490$&$1.553$\\
set$_7$&$15.500$&$18.398$&$0.701$&$6.193$&$27.715$&$1.361$\\
set$_{semi}$&$15.627$&$18.329$&$0.724$&$5.327$&$26.043$&$1.099$\\ \mr
average&$15.642$&$18.650$&$0.717$&$6.293$&$28.797$&$1.458$\\
$\sigma$&$0.138$&$0.383$&$0.012$&$0.801$&$1.687$&$0.175$\\
$\%$ var & 0.9        & 2.0        &  1.7     &  12.7   &    5.8       &    12.0 \\
\hline\hline \end{tabular} \end{center}
\end{table}

\subsection{Analysis of the LDM3 formula}

Here we show the results obtained using Eq. (\ref{roger})  for the nine regions.
In our calculations we select nine sets of fixed values for the parameters $a_v$, $a_s$, $r_0$, $a_{pair}$, $k_{v}$, $k_s$, $f_p$, $a_{c,exc}$, $a_k$ and$ k_k$. The results are exhibited in Tables \ref{tab5} and \ref{tab6}.

\begin{table}[h]
\begin{center}
\caption {RMS (in keV) for the nine groups (columns), for the nine sets of parameters (rows), employing Eq. (\ref{roger})}
\label{tab5}
\bigskip
\begin{tabular}{c|c|ccccccc|c}
\hline \hline
       & all  &	1	&2	&3	&4	&5	&6	&7	&semi-magic\\\hline
set$_{all}$&${\bf 2422}$&$1884$&$3773$&$3186$&$2031$&$1519$&$1861$&$2125$&$3937$\\ \mr
set$_1$&$3576$&${\bf 1183}$&$3067$&$3785$&$3461$&$3456$&$5854$&$1283$&$4424$\\
set$_2$&$4211$&$3032$&${\bf 1597}$&$2648$&$2928$&$4409$&$6596$&$4871$&$2333$\\
set$_3$&$3563$&$3398$&$2197$&${\bf 2151}$&$2618$&$3628$&$4455$&$4808$&$2499$\\
set$_4$&$3049$&$3035$&$3849$&$4090$&${\bf 1517}$&$1764$&$2560$&$3456$&$3952$\\
set$_5$&$2776$&$2406$&$4838$&$4004$&$1927$&${\bf 986}$&$1587$&$2110$&$4464$\\
set$_6$&$2895$&$2083$&$5073$&$4459$&$2701$&$1471$&${\bf 819}$&$1579$&$5103$\\
set$_7$&$3200$&$1484$&$4128$&$4317$&$3506$&$2701$&$3790$&${\bf 629}$&$5135$\\ \mr
set$_{semi}$&$4707$&$3776$&$1781$&$2720$&$3337$&$4962$&$7321$&$5392$&${\bf 1967}$\\
\hline\hline \end{tabular} \end{center}
\end{table}

\begin{table}[h]
\begin{center}
\caption {Sets of parameters which minimize the RMS for the nine groups of nuclei, employing Eq. (\ref{roger}). In the last three rows the average value of each parameter, their dispersion and percentage variation $\% \hbox{var}=100 \frac {\sigma}{\mid \hbox{average} \mid }$  are shown.}
\label{tab6}
\bigskip
\begin{tabular}{r|rrrrrrrrrr}
\hline \hline
&$a_v$&$a_s$&$r_0$&$a_{pair}$&$k_{v}$&$k_s$&$f_p$&$a_{c,exc}$&$a_k$&$k_k$\\
\hline
set$_{all}$&$15.647$&$20.610$&$1.198$&$-1.002$&$2.079$&$4.451$&$1.923$&$0.296$&$-4.738$&$30.717$\\
set$_1$&$15.834$&$21.074$&$1.203$&$-0.905$&$1.305$&$-0.812$&$1.265$&$0.304$&$-5.910$&$-12.666$\\
set$_2$&$15.557$&$20.860$&$1.213$&$-1.037$&$1.954$&$4.339$&$2.583$&$0.418$&$-4.719$&$32.562$\\
set$_3$&$15.673$&$19.718$&$1.234$&$-1.005$&$1.734$&$2.072$&$2.071$&$0.394$&$-1.248$&$21.077$\\
set$_4$&$15.671$&$22.448$&$1.227$&$-0.726$&$2.720$&$9.059$&$2.936$&$0.661$&$-7.069$&$58.552$\\
set$_5$&$15.577$&$21.060$&$1.196$&$-0.873$&$2.398$&$6.963$&$2.259$&$0.334$&$-6.359$&$44.965$\\
set$_6$&$15.611$&$20.997$&$1.158$&$-0.963$&$2.403$&$6.781$&$1.457$&$0.084$&$-8.514$&$33.205$\\
set$_7$&$15.806$&$20.834$&$1.204$&$-0.962$&$1.393$&$-0.390$&$1.393$&$0.289$&$-4.988$&$-12.233$\\
set$_{semi}$&$15.363$&$18.242$&$1.221$&$-0.904$&$1.418$&$0.666$&$2.577$&$0.211$&$0.538$&$-0.408$\\ \mr
average&$15.638$&$20.649$&$1.206$&$-0.931$&$1.934$&$3.681$&$2.052$&$0.332$&$-4.779$&$21.752$\\
$\sigma$&$0.132$&$1.077$&$0.021$&$0.089$&$0.479$&$3.313$&$0.558$&$0.149$&$2.662$&$23.694$\\
$\%$ var & 0.8        & 5.2        &  1.7     &  9.5   &    24.8       &    90.0  & 27.2  &   44.9  & 55.7 & 108.7
 \\\hline\hline \end{tabular} \end{center}
\end{table}

\subsection{LDM and deformation}

Tables \ref{tab1}, \ref{tab3}, \ref{tab5} display the RMS obtained with the three LDM formulas. Each row refers to one fixed set of parameters, each column to one group of nuclei. 

For the LDM1, Table \ref{tab1}, the global RMS is 2.39 MeV for the 2149 nuclei. In some groups, containing around two or three hundred nuclei, the RMS obtained with this set of parameters is smaller than the global one. The largest RMS are found in groups 2 and 3, those containing spherical nuclei.
Along each column, corresponding to one group of nuclei, the smallest RMS always corresponds the set of parameters obtained fitting in this group, as expected. These RMS  values are displayed in bold numbers. It is remarkable that the smallest RMS values are found in the two groups having the more oblate deformed nuclei, in groups 6 and 7, with 0.87 and 0.75 MeV, respectively. On the  other hand, the best fit of the spherical nuclei in group 3 has an RMS larger than 2.0 MeV, and those nuclei with very small quadrupole deformation, belonging to groups 2 and 4, have RMS larger than 1.6 MeV. Consistently, the semi-magic group of nuclei, which have small quadrupole deformations, have an RMS of 2.1 MeV.

By analyzing the RMS for each column, it is possible to notice that the smaller RMS are always found around the smallest one. It supports the idea that the division in the seven groups with different deformations makes sense, because the parameters obtained fitting nuclei with a close deformation produce also a small RMS.  Notice, for example, that among the RMS of groups 1 and 7, the more oblate and prolate deformed, respectively, the largest RMS are found with the sets of parameters  2 and 3, i.e. those fitted for spherical nuclei. On the opposite side, along columns 3 and 4, containing the more spherical nuclei, the largest RMS are found employing the sets 1 and 7, and in some cases 2 and 6.

It is tempting to conclude that the LDM is best suited for the description of prolate quadrupole deformed nuclei. In order to find support for this conclusion, it is worth to analyze the results presented in Tables \ref{tab3} and \ref{tab5}.  The global RMS for LDM2 is  2.37 MeV, and for LDM3 is 2.42 MeV. The three fits are pretty close to each other. The smallest RMS are always found for the 364 nuclei belonging to group 7, the more prolate deformed, with 0.65 MeV  and 0.63 MeV for LDM2 and LDM3, respectively.
The masses of the 332 spherical  nuclei included in group 3, and the 185 semi-magic nuclei can hardly be fitted with and RMS smaller that 2.0 MeV. Also the correlations between groups are similar for these two other models. 

This is the most relevant result reported in this contribution: {\bf the Liquid Drop Model is best suited to describe the masses of prolate deformed nuclei than of spherical nuclei}. 

\subsection{Comparison of the  three LDM formulas}

While being close to each other, the three LDM formulas employed in this work have differences which are worth to be studied in detail.  A statistical analysis of the parameters of each  model probed to be useful in previous studies \cite{Men08}. From the nine values of each parameter, their average, fluctuations and percentage fluctuations are presented in the last three rows of Tables \ref{tab2}, \ref{tab4} and \ref{tab6}.
 
The LDM1, Eq. (\ref{piet}) and LDM2, Eq. (\ref{wang}) are very similar. Their volume, surface and Coulomb parameters, listed in Tables \ref{tab2} and \ref{tab4}, have nearly equal numerical values, and are all of them very stable, with a dispersion along the nine sets smaller than 2\%.  The pairing parameter fluctuate between 10\% and 13\%, and the coefficient of the symmetry term around 6-7\%. The surface symmetry term $y$ in LDM1 is the most unstable, with fluctuations of 22\%, while its counterpart in LDM2, $\kappa$, has fluctuations of 12\%. In this subtle sense, LDM2 could be considered more stable than LDM1. It could be useful to perform a deeper comparison of the surface symmetry terms, which in one model was obtained asking for consistency between the volume and surface contributions, while in the other was designed to incorporate the isospin dependence explicitly.

The LDM3, Eq. (\ref{roger}),  has  also stable volume and Coulomb (charge radius) terms, but the surface term has fluctuations of the order of 5\%, as seen in Table \ref{tab6}. The pairing parameter, as in the other two models, fluctuates around 10\%. On the other hand, the model has ten parameters, and the remaining six have enormous fluctuations in both magnitude and sign. These instabilities of the model parameters could be interpreted as a weakness which should be addressed. 

\section{Shell effects}

The main obstacle for an accurate description of spherical nuclei employing a Liquid Drop Mass formula are the shell effects around closed shells. 

In the literature many different ways of implementing shell corrections to the LDM can be found; in general, these methods are rather laborious. A simple method was proposed in Refs. \cite{Die07,JMT}  based on counting the number of valence nucleons. This shell correction is linear and quadratic in the total number of valence nucleons $n$ and $z$,
\begin{equation}
E_{LDMM} = E_{LDM1} + b_1 (n+z) + b_2  (n+z)^2 
\label{shell}
\end{equation}
where $n$ and $z$ are the numbers of valence neutrons and protons (particle- or hole-like) and $b_i$ are parameters. Inclusion of these two terms in the LDM mass formula (\ref{piet}) reduces the rms deviation from 2.39 to 1.05MeV.

Following Ref. \cite{Die09}, we employ also an upgraded version of the terms (\ref{shell}), which is suggested by the microscopic mass formula of Duflo and Zuker \cite{Duf95,Zuk08}:
\begin{equation}
E_{LDMM'} = E_{LDMM} + a_1 S_2 + a_2 (S_2)^2 + a_3 S_3 + a_{np} \, S_{np} 
\label{ldmmp}
\end{equation}
where
\begin{eqnarray}
S_2 =\frac {n \bar n} {D_n} + \frac {z \bar z} {D_z} ,\quad
S_3  =\frac {n \bar n (n -\bar n)} {D_n} + \frac {z \bar z (z -\bar z)} {D_z} ,\quad
S_{np }= \frac {n \bar n} {D_n} \frac {z \bar z} {D_z} , 
\end{eqnarray}
with $\bar n = D_n - n$ and $\bar z = D_z - z$, where $D_n (D_z)$ is
the degeneracy of the neutron (proton) valence shell. 

They include 2-, 3- and 4-body terms. The quadratic term is associated to configuration mixing and the cubic one to a genuine three body force \cite{Men10}.

In Table \ref{tabshell} the RMS of the best fits for all the nuclei, for the seven sets of nuclei grouped according to their deformations, and for the semi-magic nuclei, are presented for the Liquid Drop Model, Eq. (\ref{piet}), for the Modified Liquid Drop Model, Eq. (\ref{shell}) and for the model including 3- and 4-body terms, Eq. (\ref{ldmmp}). The first row corresponds to the RMS listed in Table \ref{tab1} in bold face numbers. 
It is clear that the inclusion of microscopic terms improves the fits. The global RMS, for all nuclei, diminished form its LDM value of 2.39 MeV to 1.07 MeV and 0.89 MeV. The most impressive reductions in the RMS are found in the spherical nuclei grouped in region 3, which drops from 
2.06 MeV to 1.01 MeV and 0.90 MeV, and for the semi-magic nuclei, whose RMS diminishes from 2.11 MeV to 1.04 MeV and 0.82 MeV.

\begin{table}[h]
\begin{center}
\caption {RMS (in keV) of the best fit for each of the nine groups (columns), employing Eqs. (\ref{piet}), (\ref{shell}) and (\ref{ldmmp}), respectively.}
\label{tabshell}
\bigskip
\begin{tabular}{c|c|ccccccc|c}
\hline \hline
       & all  &	1	&2	&3	&4	&5	&6	&7	&semi-magic\\
\hline
LDM1&$2387$&$1313$&$1676$&$2063$&$1746$&$1053$&$869$&$746$&$2113$\\
LDMM&$1075$&$797$&$962$&$1007$&$828$&$711$&$792$&$616$&$1038$\\
LDMM'&$888$&$623$&$741$&$902$&$634$&$562$&$620$&$575$&$817$\\
\hline\hline \end{tabular} \end{center}
\end{table}

On the other hand, the nuclei in region 7, the most prolate deformed, are the best fitted in the LDMM, while the spherical nuclei in region 3 and the semi-magic nuclei have the largest RMS. The inclusion of 2-, 3- and 4-body terms in LDMM' seems to succeed in introducing deformation effects. Regions  1 and 4 to 7 have all RMS between 562 and 623 keV. Spherical and semi-magic nuclei remain to be those with the largest RMS.

\section{Conclusions}

Along this contribution we have shown that the Liquid Drop Model is best suited to describe the masses of prolate deformed nuclei than of spherical nuclei. The analysis was performed employing three different Liquid Drop Mass formulas. With them, the are nuclear masses nuclei grouped in eight sets with similar quadrupole deformations were fitted. For the three LDM models it was found that the masses of prolate deformed nuclei can be described with remarkable precision for a LDM,  with an RMS smaller than 750 keV, while the masses of spherical and semi-magic  nuclei are those worst described, with RMS larger than 2000 keV.

The dispersion of the parameters of the three models were studied comparing the fits for the different groups of nuclei. We found that in the three the surface symmetry term is the one which varies the most from one group of nuclei to another. In the model of Ref. \cite{Roger10}, isospin dependent terms were found to exhibit strong changes, making this model the least robust of the three under this criterion. 

The inclusion of shell effects allows for better fits,  which continue to be better in the prolate deformed nuclei region.  The Duflo-Zuker model is based in a microscopic description of shell effects, and describes deformation through a change in valence occupations. It remains a challenge to see if the DZ mechanism to incorporate deformation effects can be successfully employed by other models.

\section{Acknowledgements}

This work was supported in part by Conacyt, M\'exico, by FONCICYT project 94142, and by DGAPA, UNAM.

\end{document}